\documentclass[12pt,preprint]{aastex}

\shorttitle{CN, O and Na in Messier 5}
\shortauthors{Smith et al.}

\begin{document}

\title{The Pattern of CN, O, and Na Inhomogeneities on the Red Giant Branch 
of Messier 5} 

\author{Graeme H. Smith}
\affil{University of California Observatories, Lick Observatory, Department 
 of Astronomy \& Astrophysics, UC Santa Cruz, 1156 High St., Santa Cruz, CA 
 95064, USA}
\email{graeme@ucolick.org}

\author{Payal N. Modi}
\affil{The Harker School, 500 Saratoga Ave, San Jose, CA 95129, USA}
\email{payalmodi14@gmail.com}

\author{Katherine Hamren}
\affil{Department of Astronomy \& Astrophysics, UC Santa Cruz, 1156 High St., 
Santa Cruz, CA 95064, USA}
\email{khamren@ucolick.org}

\begin{abstract}
Data from the literature are used to explore the relation between $\lambda$3883
CN band strength and the sodium and oxygen abundances of red giants in the 
globular cluster Messier 5. Although there is a broad tendency for CN-strong
giants in this cluster to have higher sodium abundances and lower oxygen
abundances than CN-weak giants of comparable absolute magnitude there are
some secondary features in these relations. The oxygen abundance [O/Fe]
shows a greater range (0.6-0.7 dex) among the CN-strong giants than the 
CN-weak giants ($\approx 0.3$ dex). By contrast [Na/Fe] shows a 0.6-0.7 dex 
range among the CN-weak giants, but a more limited range of 0.3-0.4 dex among
the CN-strong giants. The $\lambda$3883 CN band anticorrelates in strength 
with [O/Fe] among the CN-strong giants, but there is little, if any, such 
trend among the CN-weak giants. In contrast, the CN band strength may show a 
modest correlation with [Na/Fe] among the CN-weak giants, but there is little 
evidence for such among the CN-strong giants. Neither oxygen or sodium 
abundance define a continuous relation with CN band strength. Instead, the 
CN-strong and CN-weak giants overlap in their sodium and possibly their oxygen 
abundances. At oxygen abundances of ${\rm [O/Fe]} = 0.20 \pm 0.05$ it 
is possible to have both CN-weak and CN-strong giants, although there may be
a discontinuity in [O/Fe] between these two groups of stars that has been
smeared out by observational errors. Both CN-weak and CN-strong giants populate
the sodium abundance range $0.4 \leq {\rm [Na/Fe]} \leq 0.6$. Messier 5
may be displaying the results of spatially heterogeneous chemical
self-enrichment.
\end{abstract}


\section{Introduction}

The earliest studies of abundance inhomogeneities in globular clusters 
(GCs) centered around the behavior of absorption bands of CN and CH in the 
spectra of red giant branch (RGB) and asymptotic giant branch (AGB) stars.
Stars occupying near-identical places in the color-magnitude diagram of a
globular cluster can display very different strengths in the CN absorption 
bands at 3883 \AA\ and 4215 \AA\ and/or the 4300 \AA\ G-band feature that 
is very sensitive to CH absorption (Zinn 1973a,b; Norris \& Zinn 1977; 
Dickens et al. 1979; Norris \& Freeman 1979; Norris 1981; Suntzeff 1980;
Smith \& Norris 1982; Briley et al. 1992; Briley 1997). Globular clusters of 
both the halo and disk/bulge populations show such phenomena. Studies during 
the 1980's revealed that abundance inhomogeneities within GCs extend to the 
elements O, Na, and Al (e.g., Cohen 1978; Peterson 1980; Norris et al. 1981; 
Leep et al. 1986), and during the 1990's the Lick-Texas group studied the 
anticorrelations between O and Na in the northern globular clusters M15, M92, 
M3, M13, M5 and M71 (Sneden et al. 1992, 1994; Kraft et al. 1992, 1993; 
Shetrone 1996). Whereas early efforts concentrated on CN and CH, the use of 
8-10 m class telescopes since the mid-1990's has seen the emphasis in GC 
inhomogeneity studies shift to elements ranging in atomic number from O to Al 
(e.g., Kraft et al. 1997; Sneden et al. 1997, 2004; Ivans et al. 1999; 
Ram\'{i}rez \& Cohen 2002, 2003; Cohen \& Melendez 2005; Johnson et al. 2005; 
Yong \& Grundahl 2008; Carretta et al. 2006; 2007; 2009; 2010; 2011; 2012, 
2013; Gratton et al. 2007, 2013).

One particular curiosity is that whereas the CN distribution within many GCs 
is bimodal (e.g., Norris 1987; Kayser et al. 2008; Smolinski et al. 2011)
the elements in the O-Al range tend to show much more uniform spreads in
abundance, as can be seen for example in Figure 7 of the large study by 
Carretta et al. (2009). In general, the CN-strong stars in bimodal-CN GCs have 
been found to have enhanced Na and Al abundances and depleted O abundances 
relative to CN-weak stars (some of the earliest studies of this trend include 
Cottrell \& Da Costa 1981; Norris \& Smith 1983; Norris \& Pilachowski 1985; 
Lehnert et al. 1991; Brown \& Wallerstein 1992; and Drake et al. 1992). 
However, the detailed relationships between the abundances of C and N, on one 
hand, and O, Na and Al, on the other, have been less well studied than the 
relationships among the O-Al elements. 

One of the earliest globular clusters in which CN-enhanced red giants were
identified is Messier 5 (Osborn 1971; Hesser et al. 1977) based on DDO
photometry. Among RGB stars the cluster exhibits a bimodal $\lambda$3883 CN 
distribution (Smith \& Norris 1983), but a fairly uniform dispersion in the 
abundances of O and Na (e.g., Carretta et al. 2009). There are a number of 
data sets in the literature on CN absorption as well as [O/Fe] and [Na/Fe]
abundances among evolved stars in Messier 5. These data are used in the present
paper to document the relationship between CN band strengths and O and Na 
abundances of RGB and AGB stars in the cluster. This work builds on an earlier
study of Ivans et al. (2001). A detailed investigation of Na and O abundance 
trends among red horizontal branch stars in Messier 5 has been published by 
Gratton et al. (2013).

\section{The CN Index data}

In red giants of Messier 5 the most prominent CN absorption band in the
optical spectrum has a bandhead that is located near 3883 \AA, although
the 4215 \AA\ CN band can also be discerned. Consequently the
sources of CN information that are used here comprise measurements of the
strength of the $\lambda$3883 band. One of the earliest studies of CN 
inhomogeneities in M5 is that of Zinn (1977), who did not make quantitative
measurements but rather classified the appearance of the $\lambda$3883 and 
$\lambda$4215 CN bands as either normal or strong based upon visual inspection
of spectrograms obtained with the KPNO 2.1 m telescope. His work revealed
the presence of $\lambda$3883 CN inhomogeneities in Messier 5.

Spectroscopic indices that quantify the strength of the $\lambda$3883 CN band 
relative to nearby comparison regions of the spectrum have been compiled from 
the following sources: Smith \& Norris (1983, 1993), Briley et al. (1992), 
Briley \& Smith (1993), Smith et al. (1997), and Langer et al. (1985, 1992).
In all but the last two of these papers measurements are given of an index 
denoted $S(3839)$ that was originally introduced by Norris et al. (1981).
It is defined in the form $S(3839) = -2.5 \log_{10} F_{\rm CN}/F_{\rm comp}$,
where $F_{\rm CN}$ is the integrated flux or intensity in the spectrum 
of a red giant across the wavelength range 3846-3883 \AA\ containing the
CN absoption feature, while $F_{\rm comp}$ is the integrated flux or intensity
in a comparison region 3883-3916 \AA\ that is reasonably free of CN absorption.
Langer et al. (1985) presented two different $\lambda$3883 CN indices 
which they designated as $m_{\rm CN}$ and $m_{\rm CN}(3883)$. Of 
these it is their $m_{\rm CN}$ measurements that have been used here because
this index is defined in a similar manner to $S(3839)$, except that the CN 
feature and comparison wavelength ranges are 3850-3878 \AA\ and 3896-3912 \AA\
respectively. Langer et al. (1992) measured an index denoted by them as 
$m({\rm CN})$ that compared the $\lambda$3883 CN absorption 
in the wavelength range 3850-3885 \AA\ to a combination of flux in the 
comparison intervals 3650-3780 \AA\ and 4020-4130 \AA.

Thus CN indices have been obtained from seven different literature sources. It 
is necessary to check for zero-point differences in the index systems of these 
various papers. We have sought to place the indices from other papers onto the 
system of Smith \& Norris (1983; SN83) simply because of their larger sample.
The Smith \& Norris (1993; SN93) $S(3839)$ indices (from their Table 1) were 
already transformed onto the SN83 system. Where there are stars in common 
between SN83 and SN93 the latter $S(3839)$ values are to be preferred since 
they have MMT or AAT measurements averaged with the SN83 values. Further 
adopted are the relations $S(3839)$[SN83] = $S(3839)$[Briley \& Smith 1993] and
$S(3839)$[SN83] = 0.1 + $S(3839)$[Smith et al. 1997] following a discussion
in Smith et al. (1997). There are 6 stars in common between the data sets of 
SN83, Briley et al. (1992; B92), and Langer et al. (1985; LKF85) that have been
used to assess the relations between their CN index systems. These are the 
stars designated by Arp (1955) as I-50, II-50, III-52, III-59, IV-4 and IV-36.
Excluding star II-50, for which there is a large difference of 0.16, the mean 
value of the difference $S(3839){\rm [SN83]} -  S(3839){\rm [B92]}$ is 
$-0.0014$, and for the purposes of this paper it is assumed that the $S(3839)$ 
indices from B92 are on the same system as SN83. Comparing the $S(3839)$ 
indices of SN83 with the $m_{\rm CN}$ data of LKF85, and again excluding II-50,
there is an average difference of 
$\langle S(3839){\rm [SN83]} - m_{\rm CN}{\rm [LKF85]} \rangle = 0.006$. This 
offset is used to transform the CN indices of LKF85 onto the system of SN83. 
Converting the $m({\rm CN})$ indices of Langer et al. (1992; LSK92) is based on
one star, Arp II-85, that is common to the study of Briley \& Smith (1993; 
BS93). For this star $S(3839){\rm [BS93]} - m({\rm CN}){\rm [LSK92]} = -1.298$.
Since the $S(3839)$ indices of BS93 are taken to be on the same scale as
SN83, the Langer et al. (1992) values of $m({\rm CN})$ were converted to the 
SN83 system of $S(3839)$ via this offset of $-1.298$. Upon transforming all CN 
indices onto the $S(3839)$ system of Smith \& Norris (1983) equal weight was 
given to all measurements in forming average values of $S(3839)$, with the 
exception noted above that where $S(3839)$ data were available from both the
SN93 and SN83 papers the former was chosen in place of the latter.
 
The CN data sources employed here typically identified the stars in their 
observing programs according to the nomenclature in the color-magnitude diagram
study of Arp (1955). Consequently the Arp designations are used in this paper.
The $V$ and $B-V$ photometry for stars with Arp (1955) designations was taken
from Sandquist \& Bolte (2004). In the case of stars S344 and S445 the 
photometry adopted is that tabulated by Smith \& Norris (1993) based on 
unpublished values from M.~Simoda. The reddening and apparent distance modulus
was adopted from the 2003 version of the catalog of globular cluster properties
described by Harris (1996): $E(B-V) = 0.03$ and $(m-M)_V = 14.46$.

Table 1 contains the data that have been compiled for this paper. Star
designations given in column 1 are those of Arp (1955).  Absolute visual
magnitudes and dereddened $(B-V)_0$ colors based on the photometry of
Sandquist \& Bolte (2004) are given in columns 2 and 3, except for stars
S344 and S445 as discussed above. Column 4 contains a list of the merged
values of the $S(3839)$ $\lambda$3883 CN index. 

There are eleven stars in Table 1 for which the listed value of $S(3839)$ is
based on measurements taken from two or three literature sources. Such 
instances are denoted with a footnote in the table. For small samples of 2 or 3
measurements the ratio of the range $R$ to the standard deviation $\sigma$ in 
a measured quantity can be taken as 1.128 and 1.693 respectively (Snedecor 
1946; Montgomery 1996). An estimate of the uncertainty representative 
of the $S(3839)$ data for stars with multiple measurements was calculated
from these $R/\sigma$ ratios and the range in the individual transformed 
$S(3839)$ values. The mean value thereby derived for $\sigma[S(3839)]$ is 0.03.
Uncertainties in the measurements of $S(3839)$ from the published studies of 
Messier 5 are typically found to be $\sigma \sim$ 0.02-0.06 mag (Smith \& 
Norris 1983; Briley et al. 1992; Briley \& Smith 1983). Various comparisons 
studied by Langer et al. (1992) indicate that 
$\sigma[m({\rm CN})] \sim 0.03$-0.04.

Over much of the magnitude range of interest the red giant branch and the 
asymptotic giant branch are well separated in the color-magnitude diagram (CMD)
of $M_V$ versus $(B-V)_0$ shown in Figure 1. There are 46 stars with $S(3839)$ 
index values in Table 1 of which 10 are AGB stars on the basis of their 
position in Fig.~1. Stars considered to be in the RGB and AGB phases of
evolution are shown as filled squares and open squares respectively. The 
classification of each star is listed in column 8 of Table 1.

The behavior of the CN index $S(3839)$ versus $M_V$ is shown in Figure 2.
At any given magnitude there is considerable scatter in CN index among the
RGB stars, which are depicted with filled and open circles according to
whether the $\lambda$3883 CN band is considered to be strong or weak
respectively. As first noted in Smith \& Norris (1983) the CN distribution
on the red giant branch of Messier 5 is bimodal, in fact, this cluster is
one of the archetype examples of CN-bimodality among Galactic globular
clusters. There is one star in Fig.~2 that is represented by an
eight-pointed symbol rather than a circle, this RGB star is IV-34
and may be an example of a red giant with intermediate CN band strength.
In addition to the scatter at a given absolute magnitude, both
the CN-weak and CN-strong stars exhibit a behavior of increasing mean 
$S(3839)$ with increasing luminosity on the RGB, which can be attributed at
least in part to the sensitivity of the CN band strength to photospheric
effective temperature. Based on Fig.~2 a reasonable lower envelope to the 
RGB-star data is one that passes through the points $(S(3839), M_V) = (0,0)$ 
and (0.2, $-2.0$). Thus a suitable baseline in Fig.~2 above which most
RGB stars fall has an equation $S(3839)_{\rm b} = -0.10 M_V$. 
Residuals $\Delta S(3839) = S(3839) - S(3839)_{\rm b}$ were calculated 
relative to this baseline for all stars listed in Table 1. The
$\Delta S(3839)$ index is an attempt to provide a measure of $\lambda$3883
CN band strength that has been compensated to at least first order for the
differences in effective temperature among the RGB stars in the sample. This
type of empirical correction has been used often in the literature going
back to studies such as Norris et al. (1981). Values of $\Delta S(3839)$
are tabulated in column 5 of Table 1 for both RGB and AGB stars, although it 
must be stressed that the baseline used to produce these residuals is defined 
largely by RGB stars.

Asymptotic giant branch stars are plotted in Fig.~2 as either triangles or
three-pointed symbols. Two AGB stars (filled triangles) fall among the 
CN-strong RGB stars in Fig.~2, while one AGB star (open triangle) sits among
the CN-weak RGB stars. However, six of the AGB stars have $M_V \sim -1$ and 
are located at intermediate positions between the CN-weak and CN-strong RGB
sequences in Fig.~2. An interpretation that the majority of the AGB stars
should be classified as CN-intermediate is not necessarily appropriate,
however, since at $M_V \sim -1$ the AGB is considerably bluer by
0.07-0.10 mag in $(B-V)$ than the RGB (Fig.~1). As such, these AGB stars 
are hotter than RGB stars of similar magnitude, and their higher effective
temperatures would serve to diminish the CN band strength relative to that
of a RGB star of comparable luminosity. When the AGB and RGB stars are instead
compared in a plot of $S(3839)$ versus $(B-V)_0$, as shown in Figure 3 (within 
which the symbols are the same as in Fig.~2), those AGB stars seemingly
of intermediate CN strength on the basis of Fig.~2 now appear to be more 
consistent with an extrapolation of the CN-strong RGB sequence. Thus, there may
also be a bimodal CN distribution among the AGB stars in Messier 5, but the
effect is muted in a plot of CN index versus absolute magnitude.

\section{The Sodium and Oxygen Abundance Data}

Sodium and oxygen abundances were compiled from two sources: Ivans et al. 
(2001), denoted I01 in the following text, and Table 6 of Carretta et al.
(2009), henceforth designated as C09. The former source contains [Na/Fe]
and [O/Fe] measurements that were derived from high-resolution echelle
spectra acquired with either the HIRES spectrometer on the Keck I telescope
or the Hamilton spectrometer on the Shane 3 m telescope of Lick Observatory.
The C09 abundances were downloaded from the electronic database of the 
VizieR Catalog Server (Table J/A+A/505/117) maintained by the Strasbourg 
Astronomical Data Center (Ochsenbein et al. 2000). They are based on spectra 
obtained with the FLAMES/GIRAFFE high-resolution multi-fiber spectrograph 
on the VLT UT2 telescope (Pasquini et al. 2002). Together these two sources
provide a considerable overlap of sodium and oxygen abundance measurements
with the CN dataset of Table 1. Oxygen and sodium abundances for M5 giants are 
also available from Ram\'{i}rez \& Cohen (2003) but CN indices are not 
available for many of the stars in their study.

A comparison between the [O/Fe] abundances of Carretta et al. (2009) and 
Ivans et al. (2001) is shown in Figure 4. Measurements by I01 that are based 
on Keck/HIRES observations are shown as filled circles while those derived
from Lick/Hamilton spectra are depicted as open circles. If the entire sample 
of I01 abundances are considered then a modest offset between the 
C09 and I01 abundances is derived, such that the mean difference from 12 stars 
is $\Delta$[O/Fe](C09 $-$ I01) = 0.07 with a standard deviation of 
$\sigma = 0.16$ dex. By contrast, if just the Keck observations of I01 are 
considered then $\Delta$[O/Fe](C09 $-$ I01[Keck]) = 0.12 and 
$\sigma = 0.15$ dex for 8 stars, while for the I01 Lick observations alone
$\Delta$[O/Fe](C09 $-$ I01[Lick]) = $-0.04$ and
$\sigma = 0.12$ dex for 4 stars. For the purposes of obtaining a combined data 
set of oxygen abundances it is considered on the basis of these comparisons
that the Lick [O/Fe] abundances of I01 are on the same system as that of C09, 
while the Keck [O/Fe] values of I01 can be placed on the system of C09 by 
increasing them by 0.12 dex. A homogeneous set of [O/Fe] values was thereby 
compiled for stars in Table 1 on an abundance scale corresponding to that of 
Carretta et al. (2009). In the case of stars II-85 and IV-47 the I01 abundances
based on their Keck HIRES spectroscopy are adopted as opposed to their 
reanalysis of Lick 3 m Hamilton echelle spectra.

A consideration of [Na/Fe] reveals evidence for larger offsets between the
abundance scales of C09 and I01, as shown in Figure 5. The mean offset between 
the C09 and I01[Keck] measurements is
$\Delta$[Na/Fe](C09 $-$ I01[Keck]) = 0.23 with 
$\sigma = 0.09$ dex for 8 stars, while for the Lick-based abundances of
I01 the offset is $\Delta$[Na/Fe](C09 $-$ I01[Lick]) = 0.42 with 
$\sigma = 0.22$ dex for 4 stars. Thus the Keck and Lick [Na/Fe] abundances
of Ivans et al. (2001) have been transformed onto the abundance scale of
Carretta et al. (2009) by adding 0.23 dex and 0.42 dex respectively. The
systematic difference between the C09 and I01 sodium abundances partly reflects
the application of non-LTE corrections to [Na/Fe] by Carretta et al. (2009),
whereas no such corrections were employed by Ivans et al. (2001).

Carretta et al. (2009) reported star-to-star errors in their [O/Fe] and [Na/Fe]
determinations of 0.14 dex and 0.08 dex respectively, upon taking into account 
contributions due to uncertainties in effective temperature $T_{\rm eff}$, 
surface gravity, microturbulent velocity $v_t$, cluster metallicity, and 
absorption line equivalent width $EW$. They found that the main contributing 
sources of error in the oxygen and sodium abundances are errors in 
$T_{\rm eff}$, $v_t$ and $EW$.

Ivans et al. (2001) do not quote formal uncertainties in their [Na/Fe] and 
[O/Fe] determinations. One can use the numbers given in their Table 3 to 
estimate the uncertainties that would result from a combination of the typical 
errors in $T_{\rm eff}$, surface gravity, $v_t$, cluster
metallicity, distance modulus, and stellar mass. These factors contribute to 
uncertainties of 0.08 dex in both [O/Fe] and [Na/Fe]. However, such estimates 
do not take into account the errors of measurement in O or Na absorption
line equivalent widths, and so can only be considered as lower limits to the
uncertainties in the [O/Fe] and [Na/Fe] abundances from their study.   
Ivans et al. (2001) have two stars, II-85 and IV-47, for which they obtained
abundances from both Keck/HIRES and Lick/Hamilton spectra. For these
two stars they obtain [O/Fe] abundances of +0.08 and +0.02 from Keck and
+0.23 and +0.14 from Lick spectra. The [Na/Fe] abundances are +0.38 and +0.17
from Keck and +0.26 and +0.07 from Lick. The data internal to the I01 study 
are consistent with uncertainties in their [O/Fe] and [Na/Fe] measurements of 
$\sim 0.10$-0.15 dex.

Based on the transformations described in the previous two paragraphs
homogenized sets of sodium and oxygen abundances have been compiled for those 
40 stars in Table 1 that were studied by either Ivans et al. (2001) and/or 
Carretta et al. (2009). The homogenized values of [O/Fe] and [Na/Fe] are listed
in columns 6 and 7 of Table 1. Of this sample oxygen and/or sodium abundances 
were measured for 21 stars by Carretta et al. (2009), and their results are 
listed in columns 9 and 10. Six of the stars in Table 1 having either [O/Fe] 
or [Na/Fe] determinations are considered to be AGB members of Messier 5.

\section{The Relation Between CN Band Strength and Oxygen Abundance}

The $\Delta S(3839)$ CN residual is plotted in Fig.~6 versus the merged 
[O/Fe] abundance data from column 6 of Table 1, with filled and open symbols 
denoting RGB and AGB stars respectively. There is a clear difference in the
mean oxygen abundances of the CN-strong and CN-weak giants, with the former
having the lower average [O/Fe], in accord with the previous finding of Ivans 
et al. (2001). Among the CN-strong population of RGB stars 
($\Delta S(3839) \geq 0.30$) there is an anticorrelation between CN band 
strength and [O/Fe]. Thus the CN-strong giants are themselves not a 
homogeneous population but have a dispersion in both $\Delta S(3839)$ and 
[O/Fe], with the oxygen abundance extending over a range of at least 0.6 dex. 
The CN-weak RGB stars ($\Delta S(3839) \leq 0.20$) exhibit a smaller range in 
[O/Fe] of $\approx 0.35$ dex, and there is no clear CN-O anticorrelation within
Fig.~6 among the CN-weak population. What is striking from Fig.~6 is that near 
the transition in oxygen abundance between CN-strong and CN-weak stars, which 
occurs at [O/Fe] $\approx +0.2$ dex, there are a number of red giants that have
nearly the same [O/Fe] yet very different $\lambda$3883 CN band strengths. It 
is almost as if the CN-strong and CN-weak RGB stars define offset sequences in 
Fig.~6, which are displaced in $\Delta S(3839)$ by about 0.3 mag at [O/Fe] = 
0.2, and which overlap over a narrow range in [O/Fe]. The one CN-intermediate 
RGB star identified in Fig.~2, Arp IV-34 with $\Delta S(3839) = 0.21$, has 
an oxygen abundance of ${\rm [O/Fe]} = +0.13$, and falls near the 
offset region in Fig.~6.   

Six AGB stars are represented in Fig.~6. One of them is an oxygen-rich CN-weak
star, while three others fall close to the sequence defined by the CN-strong
RGB stars. Two AGB stars (I-55 and III-53) stand out as having low oxygen 
content and relatively weak CN bands for their $M_V$ magnitude. They appear to 
fall on an extrapolation of the CN-weak RGB sequence to depleted oxygen
abundances. However, as noted in Sec.~2, the relatively weak CN bands
may be a consequence of the higher effective temperature of these AGB stars
relative to their RGB counterparts of comparable absolute magnitude. Thus the
outlying positions of I-55 and III-53 in Fig.~6 may be a temperature effect.

In Figure 7 the CN residual $\Delta S(3839)$ is again plotted against oxygen 
abundance except that this plot shows only the [O/Fe] abundances from Carretta 
et al. (2009) that are listed in column 9 of Table 1. In this figure there will
be no scatter in [O/Fe] due to uncertainties in the I01-to-C09 transformations.
Nonetheless, the offset between CN-strong and CN-weak RGB stars near
[O/Fe] = 0.2 is still conspicuous in Fig.~7, and the range in [O/Fe] abundance
among the CN-weak giants is still on the order of 0.30-0.35 dex. There may be 
a modest overlap in oxygen abundance between the CN-weak and CN-strong
groups evident in Figs.~6 and 7, however this overlap of $\sim 0.1$ dex is
comparable to the observational uncertainties in the [O/Fe] determinations.
As such, the data cannot rule out the possibility that there is a discontinuity
in [O/Fe] between the CN-strong and CN-weak giants that has been smeared
out in Figs.~6 and 7 by observational errors.

Whereas in broad terms it can be concluded that the CN-strong RGB stars in 
Messier 5 have lower [O/Fe] abundances than the CN-weak red giants, the 
situation is modified by the fact that (i) there is a spread of more than 
0.6 dex among the CN-strong giants, and (ii) there is a population of red 
giants that have [O/Fe] $\sim +0.2$ but nonetheless very different 
$\lambda$3883 CN band strengths.

The empirical anticorrelation between CN band strength and oxygen abundance 
seen in Figs.~6 and 7 suggests that relative to the atmospheres of the CN-weak 
giants the material in the CN-strong stars has been subject to the 
O$\rightarrow$N cycle of hydrogen burning. If the CN-strong stars contain 
within their atmospheres material that was initially like that of the CN-weak 
giants, but which has been subjected to the O$\rightarrow$N process of hydrogen
burning, then a factor of 10 or more enhancement in nitrogen abundance might 
be found among the CN-strong giants.

The CN-weak giants in Messier 5 are worthy of additional study. It would be 
valuable to identify a larger sample of such stars and to document the range 
in [O/Fe] among them, so as to determine whether there is an anticorrelation 
between CN band strength and [O/Fe] within this population. The nitrogen 
abundances of the CN-strong giants may have been enhanced in part by a 
contribution from the C$\rightarrow$N cycle of hydrogen burning in addition to 
the action of the O$\rightarrow$N process within some initial CNO abundance 
mix. There is modest evidence of a CN-CH anticorrelation on the RGB of 
Messier 5 (Smith \& Norris 1983), however, their data are in part based on 
photographic spectra, and a modern CCD study of carbon abundances among both 
CN-weak and CN-strong giants aimed at determining the precise 
relations between C, N, and O abundances would be another worthwhile pursuit.
Cohen et al. (2002) did uncover a CN-CH anticorrelation among stars on the 
lower half of the red giant branch of Messier 5. A spectrum synthesis analysis 
is needed to determine whether the total C$+$N$+$O abundance is the same or 
different in the CN-strong and CN-weak giants.

\section{The Relation Between CN Band Strength and Sodium Abundance}

Figures 8 and 9 contain plots of the $\Delta S(3839)$ CN residual versus
[Na/Fe] abundances from columns 7 and 10 respectively of Table 1. Thus the
sodium abundances in the larger sample of Fig.~8 are from the merging of the
I01 and C09 data sets, whereas only the C09 abundances are used in Fig.~9.
Filled and open squares again denote RGB and AGB stars respectively. It is
clear from both figures that the CN-strong stars as a population have 
higher mean sodium abundances than the CN-weak stars, again consistent with
the earlier findings of Ivans et al. (2001). Most CN-strong giants 
observed in Messier 5 have sodium abundances of ${\rm [Na/Fe]} \geq +0.4$, 
whereas many of the CN-weak giants have [Na/Fe] less than 0.4. 
However, there are some additional interesting features. 

The CN-strong giants exhibit a 0.4 dex range in [Na/Fe], and
the majority of them have [Na/Fe] $\leq 0.7$. By contrast the CN-weak giants
exhibit a notably wider spread in [Na/Fe] of $\approx$ 0.7 dex. There may be
a mild trend in Fig.~8 for the $\Delta S(3839)$ residual to correlate with
[Na/Fe] among the CN-weak red giants, which is perhaps better seen in Fig.~9 
despite the smaller number of stars. Among the CN-strong RGB stars there is a 
smaller range in [Na/Fe] and no evidence of a CN-Na correlation. The sodium 
abundance range of $0.4 \leq {\rm [Na/Fe]} < 0.6$ is populated by both 
CN-strong and CN-weak giants, and it is possible to identify pairs of CN-strong
and CN-weak giants that have identical sodium abundances within the 
observational uncertainties. It appears that the 
CN-strong and CN-weak sequences in Fig.~8 overlap by 0.2 dex in [Na/Fe], but 
that they are offset by $\sim 0.2$ mag in the $\Delta S(3839)$ residual. This 
conclusion might in part be a consequence of uncertainties in the L01-C09 
transformations smearing out the [Na/Fe] values around 0.4 dex in Fig.~8. 
Nonetheless, in Fig.~9 where only the [Na/Fe] abundances from Carretta et al.
(2009) are used, two CN-weak RGB stars with ${\rm [Na/Fe]} > 0.35$ overlap 
the CN-strong giants in sodium abundance. 

\section{Discussion}

The behavior of oxygen and sodium as a function of the CN band strength 
of red giants in Messier 5 shows some interesting contrasts. Whereas [O/Fe]
shows a greater range among the CN-strong giants than the CN-weak giants,
[Na/Fe] shows a greater range among the CN-weak giants. The CN band strength 
anticorrelates with [O/Fe] among the CN-strong giants, but there is little 
tendency of such among the CN-weak giants. By contrast, although the CN band 
strength may show some modest correlation with [Na/Fe] among the CN-weak 
giants, there is little evidence for a CN-Na correlation among the CN-strong 
giants. What is common in the case of both oxygen and sodium is that there is 
not a continuous relation between CN band strength and either [O/Fe] or 
[Na/Fe]. Rather the CN-strong and CN-weak giants overlap in their range of 
both sodium and possibly oxygen abundance (sodium more so than oxygen).

On the basis of the [Na/Fe] versus [O/Fe] diagram, Carretta et al. (2009) 
divided the stars in Messier 5 into three groups. Primordial stars (or 
``first-generation'' or P stars in their terminology) were defined as those 
with ${\rm [Na/Fe]} \leq 0.1$; such stars also tend to have 
${\rm [O/Fe]} > 0.2$. Among the enriched stars (or ``second-generation'' stars 
in the terminology of C09) Carretta et al. (2009) drew a dividing line between 
an intermediate (I) and an extreme (E) component that has the equation 
${\rm [Na/Fe]} = {\rm [O/Fe]} + 0.9$. The most extreme of the enriched stars 
have ${\rm [Na/O]} > 0.9$ along with ${\rm [Na/Fe]} > 0.5$, and these represent
only a small fraction of the stars in Messier 5. The bulk of the enriched 
stars have ${\rm [O/Fe]} > -0.3$ and $0.6 > {\rm [Na/Fe]} > 0.1$. 
According to Figs.~8 and 9 the ``first-generation'' P stars of Carretta et al.
(2009) are all CN-weak. There are 5 stars in Table 1 which on the basis of
the [Na/Fe] abundances listed in column 7 would be classified in the P group
of Carretta et al. (2009). The mean and standard deviation in $\Delta S(3839)$
for these stars are 0.04 and 0.03 respectively, which is typical of the giants
with the weakest $\lambda$3883 CN bands in Messier 5. Of the 27 stars in 
Table 1 that fall into the intermediate I group the mean and standard deviation
in $\Delta S(3839)$ are 0.26 and 0.18 respectively; whereas for the extreme E 
group with ${\rm [Na/O]} > 0.9$ (on the basis of columns 6 and 7 of Table 1) 
the mean $\Delta S(3839)$ is 0.44 with a standard deviation of 0.12. Thus, the 
P, I, and E groups of Carretta et al. (2009) have progressively larger mean CN 
residuals.

What is interesting, however, is that the CN-weak population of giants with 
$\Delta S(3839) < 0.2$ does include quite a number of stars that would be 
classified in the  ``second-generation'' I group by Carretta et al. (2009) on 
the basis of having sodium abundances of ${\rm [Na/Fe]} > 0.1$, e.g., stars 
such as I-68, II-39, III-36 and IV-19. If we consider CN-weak giants to be 
those with $\Delta S(3839) \leq 0.2$ then the [Na/Fe] abundance varies from 
$-0.1$ to $0.6$ dex among these stars. Thus the sodium abundances of the 
CN-weak giants extend well into the range in [Na/Fe] typical of C09's group I. 
There are no C09 group E stars among the CN-weak giants. Among the CN-strong 
giants with $\Delta S(3839) \geq 0.3$ the lowest value of [Na/Fe] is 0.10
for star II-50, which falls right at the boundary of C09's groups P and I. 
These comparisons emphasize the point made in Section 5 concerning the overlap 
of sodium abundances between CN-strong and CN-weak giants.

The pattern of abundances in Messier 5 might be interpreted within a context 
whereby CN-strong giants formed from gas of different initial composition to 
that of the CN-weak stars. Cohen et al. (2002) showed that star-to-star CN 
band variations extend to the base of the RGB at $M_V \sim +3$ in Messier 5, 
consistent with the variations being of a very early or primordial 
origin.\footnote{On top of the complexities of primordial enrichment there 
is the added complication of interior mixing within the red giants of 
Messier 5. The dredge-up of CNO-processed material can alter the original 
composition of both CN-strong and CN-weak giants. Whereas such additional 
processing could increase the nitrogen surface abundances of both CN-weak and 
CN-strong giants, which would act to strengthen the CN bands, it would also 
reduce the surface carbon abundance, thereby having an opposing effect of 
diminishing the CN band strengths.} One suggestion on the basis of the 
CN-O anticorrelation of Fig.~6 is that the CN-strong stars might have formed 
from gas that was initially identical to that of the CN-weak giants, but was 
processed through the CNO bicycle of hydrogen burning before becoming 
incorporated into the CN-strong stars themselves. Dating back to a suggestion 
by Cottrell \& Da Costa (1981) it has become common to view the interiors of
intermediate-mass AGB stars (e.g., Fenner et al. 2004; Ventura \& D'Antona 
2008a, 2008b, 2009; Decressin et al. 2009),
or even more massive stars (e.g.,  Smith 2006; Decressin et al. 2007),
formed at early times within a globular cluster, as the sites for such 
additional CNO-processing prior to CN-strong star formation. 

Whereas the abundances of C, N, and O are the product of the CNO bicycle 
reactions of hydrogen burning, the element sodium can be manufactured in a
Ne-Na proton-addition cycle that takes place at temperatures at which the CNO 
bicycle also occurs (e.g., Langer et al. 1993; Cavallo et al. 1996).
The action of the Ne-Na cycle in bringing about a significant build up of 
sodium appears to have occurred within the confines of the CN, O, Na 
composition range prevalent among the CN-weak giants. In other words,
some stars in Messier 5 formed with an initial enrichment in
Na that was not accompanied by a great enough enhancement in nitrogen 
as to cause them to exhibit strong CN bands on the red giant branch.
Other stars may have formed with comparable sodium abundances but
high enough nitrogen as to become CN-strong red giants. Similarly,
two stars might have formed in Messier 5 with similar oxygen abundances
of [O/Fe] $\sim 0.15$-0.25 yet one with much lower nitrogen abundance
than the other, such as to lead to a CN-weak and CN-strong pair once the
stars evolved onto the red giant branch.

Messier 5 seems to be a cluster in which there is not a unique one-to-one
relationship between N abundance on one hand (as traced by CN band strength
on the RGB) and O and Na abundances on the other hand. Perhaps what 
Messier 5 is indicating is that the self-enrichment of this cluster was the 
by-product of spatially heterogeneous enrichment by intermediate-mass or 
high-mass stars having a range of ages and formation times.

This research has made use of the VizieR and SIMBAD catalog access tools, 
CDS, Strasbourg, France. We thank the referee for useful comments on the
manuscript. G.~H.~S. gratefully acknowledges support from the National
Science Foundation through grant AST-0908757.

\clearpage

\begin{figure}
\figurenum{1}
\begin{center}
\includegraphics[scale=0.7]{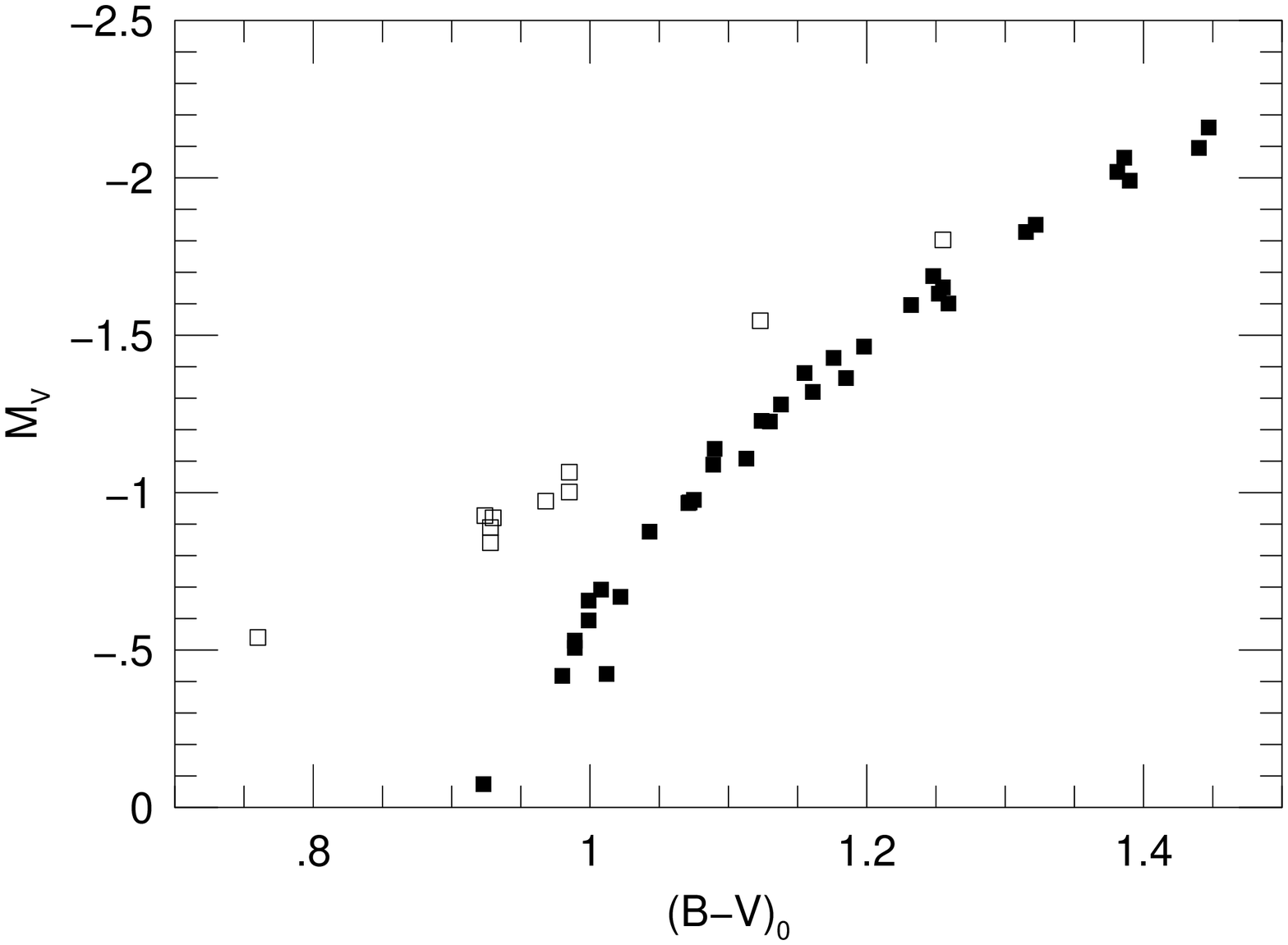}
\caption{The $M_V$ versus $(B-V)_0$ color-magnitude diagram of stars from 
 Messier 5 that are listed in Table 1. Filled symbols correspond to stars 
 considered to be on the red giant branch whereas open symbols denote 
 asymptotic giant branch stars.}
\end{center}
\end{figure}
\clearpage

\begin{figure}
\figurenum{2}
\begin{center}
\includegraphics[scale=0.7]{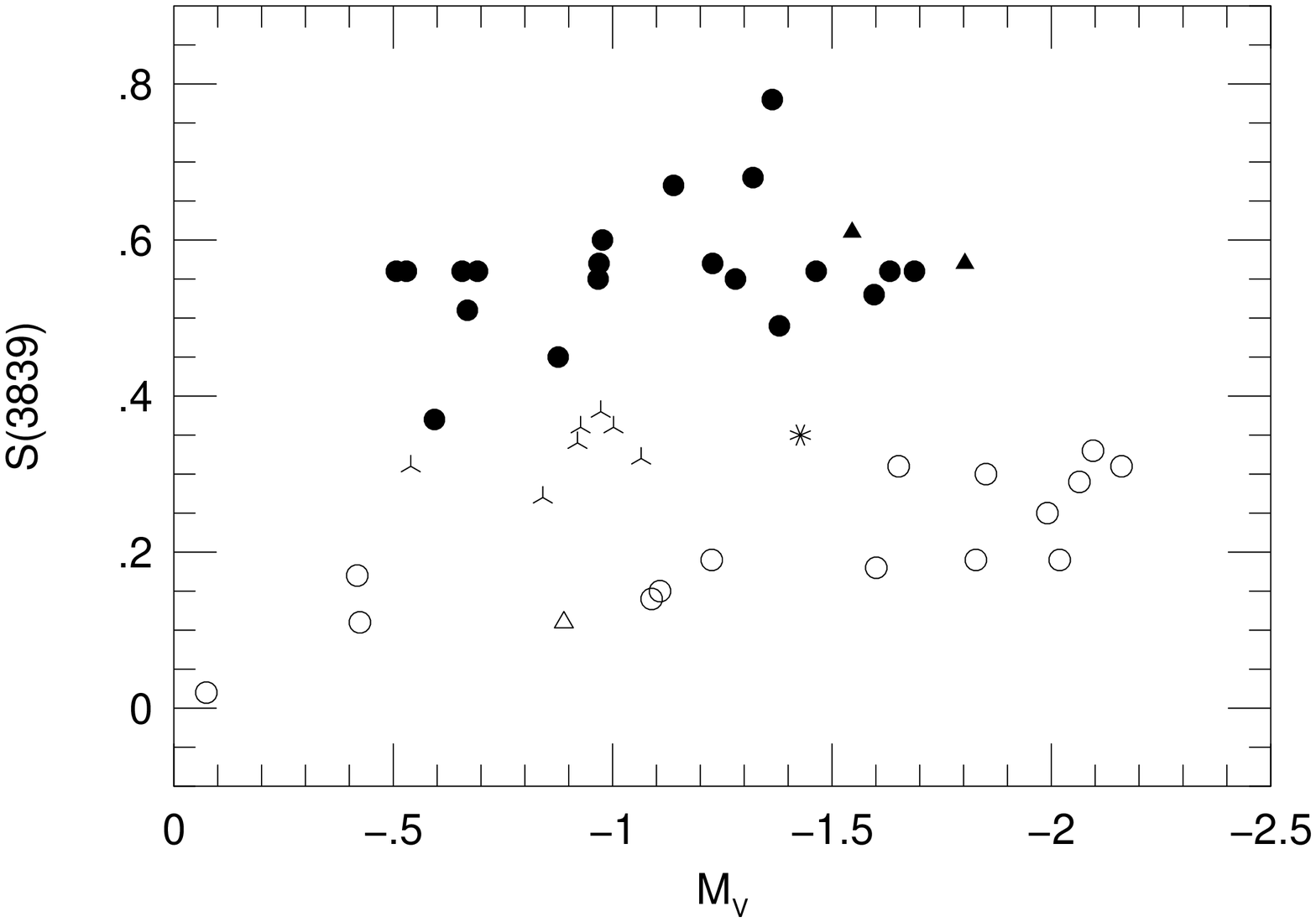}
\caption{The CN index $S(3839)$ versus $M_V$ for giant stars in Messier 5.  
 Filled and open circles correspond to CN-strong and CN-weak RGB stars
 respectively. One RGB star with intermediate CN band strength is shown as an
 eight-pointed symbol. Triangles and three-pointed symbols denote AGB stars.}
\end{center}
\end{figure}
\clearpage

\begin{figure}
\figurenum{3}
\begin{center}
\includegraphics[scale=0.7]{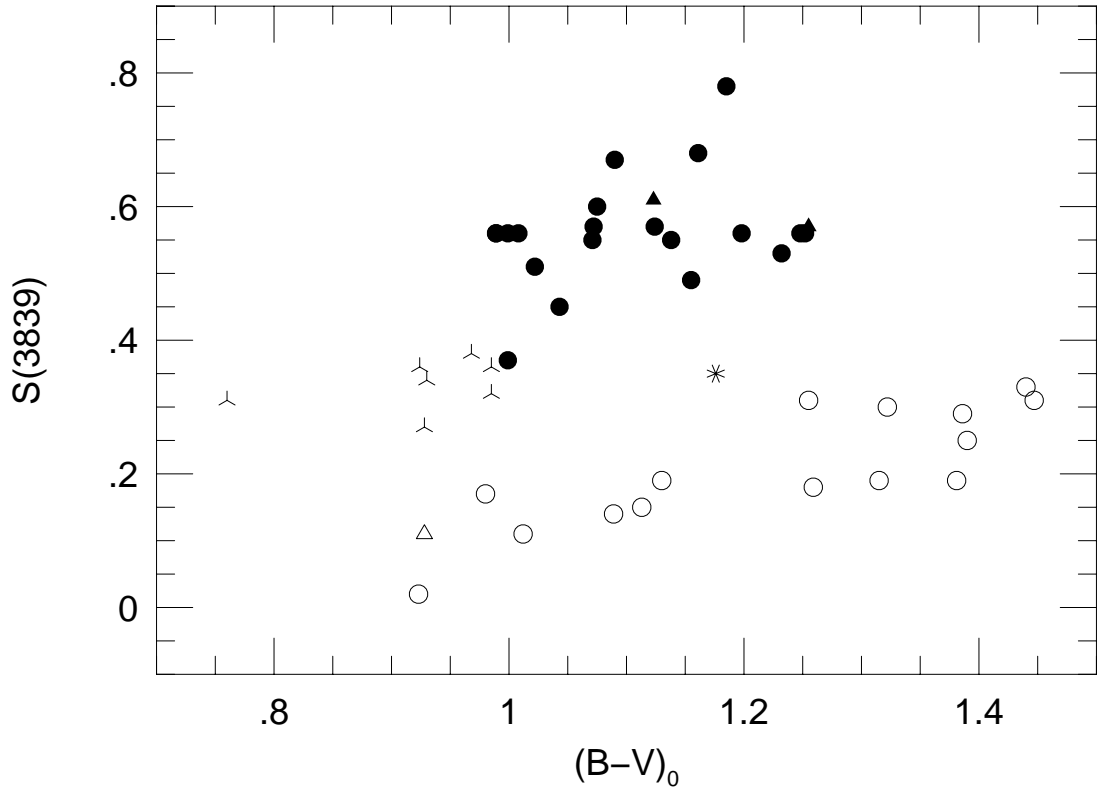}
\caption{The CN index $S(3839)$ versus $(B-V)_0$ for giants in Messier 5.  
 Filled circles, open circles, and the eight-pointed symbol correspond to 
 CN-strong, CN-weak and CN-intermediate RGB stars respectively. Triangles and 
 three-pointed symbols denote AGB stars.}
\end{center}
\end{figure}
\clearpage

\begin{figure}
\figurenum{4}
\begin{center}
\includegraphics[scale=0.7]{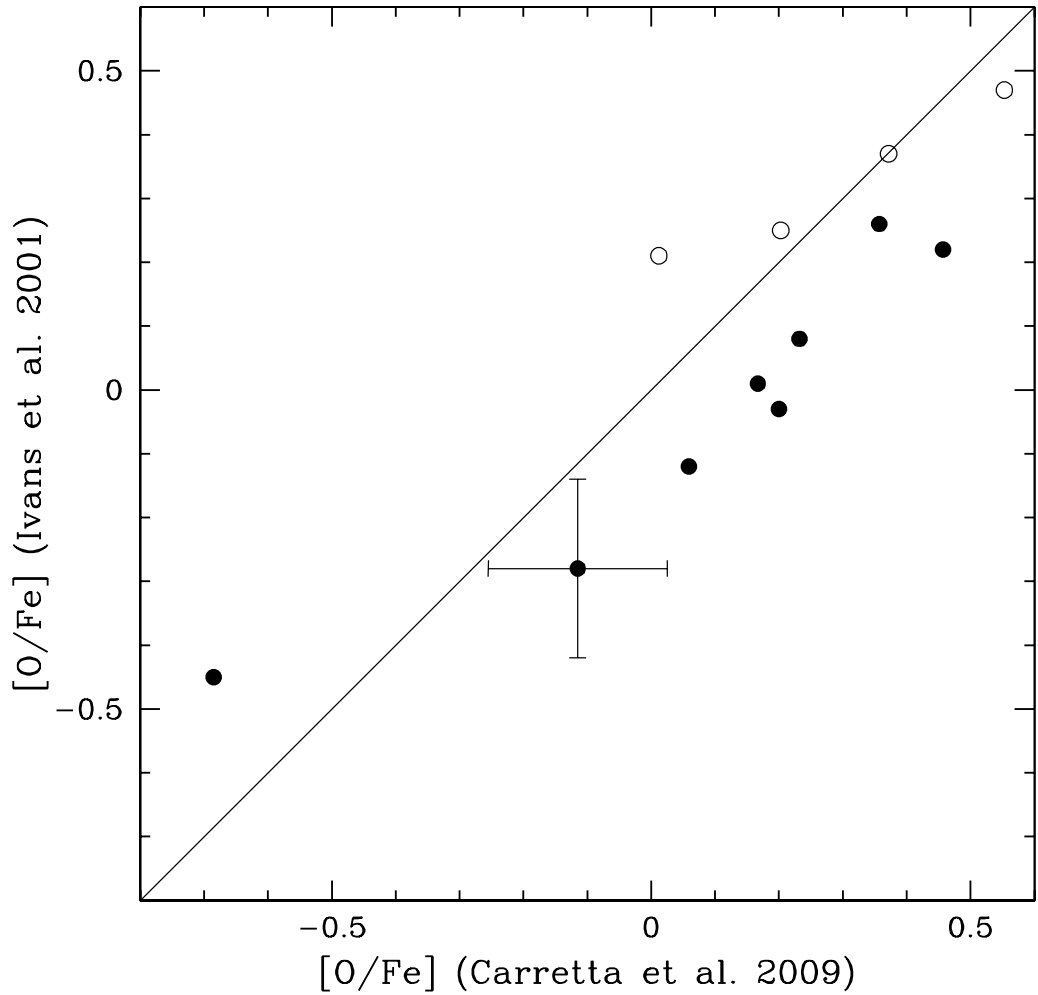}
\caption{Oxygen abundances from Ivans et al. (2001) versus those from
Carretta et al. (2009) for stars in common between the two programs. Results
from I01 based on HIRES and Hamilton spectrometer data are shown as filled
and open symbols respectively. Error bars of length $\pm 0.14$ dex in both the 
C09 and I01 [O/Fe] abundances are shown for one star.}
\end{center}
\end{figure}
\clearpage

\begin{figure}
\figurenum{5}
\begin{center}
\includegraphics[scale=0.7]{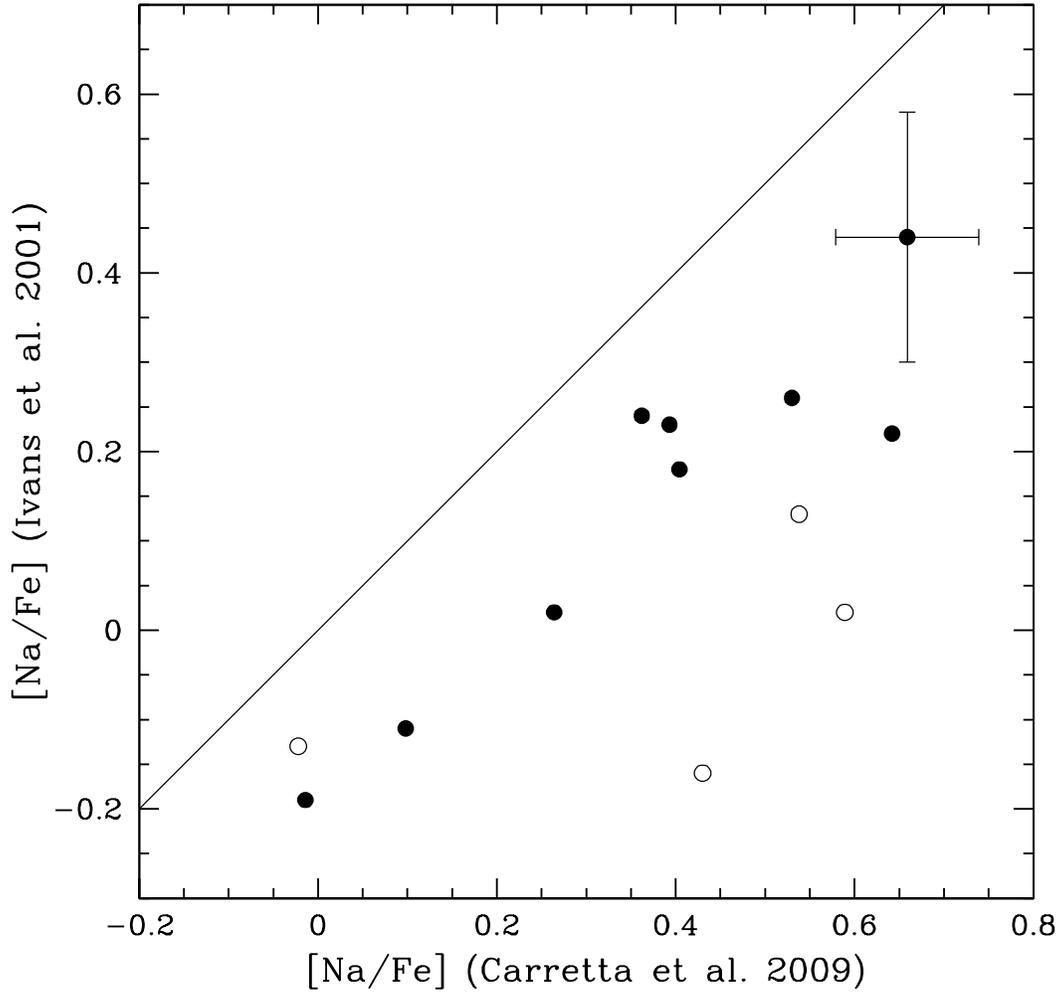}
\caption{Sodium abundances from Ivans et al. (2001) versus those from
Carretta et al. (2009) for stars in common between the two programs. Results
from I01 based on HIRES and Hamilton spectrometer data are shown as filled
and open symbols respectively. Error bars of length $\pm 0.08$ dex and 
$\pm 0.14$ dex in the C09 and I01 [Na/Fe] abundances respectively are shown 
for one star.}
\end{center}
\end{figure}
\clearpage

\begin{figure}
\figurenum{6}
\begin{center}
\includegraphics[scale=0.7]{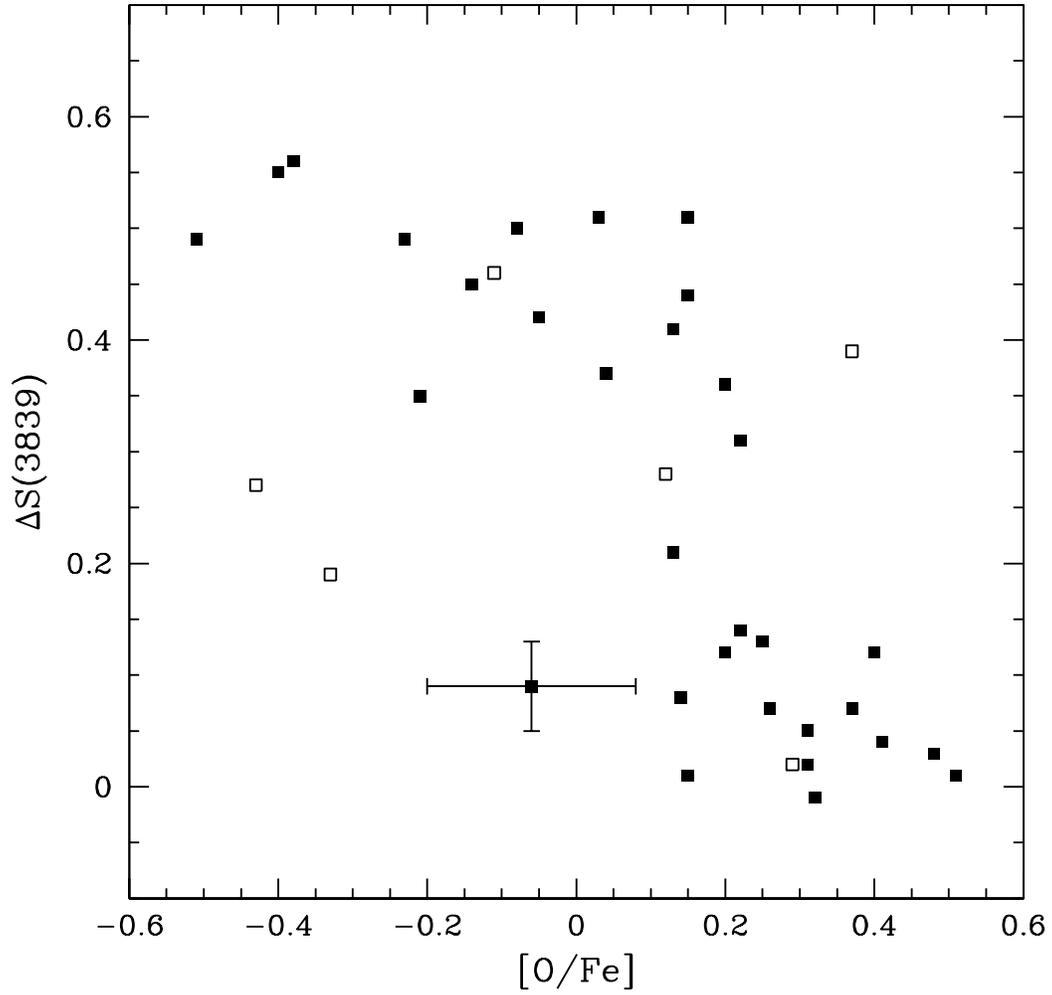}
\caption{The $\lambda$3883 CN residual $\Delta S(3839)$ versus oxygen abundance
from column 6 of Table 1. Filled and open squares denote RGB and AGB stars 
respectively (as in Fig.~1). A pair of representative error bars of length 
$\pm 0.04$ in $\Delta S(3839)$ and $\pm 0.14$ dex in [O/Fe] is depicted.}
\end{center}
\end{figure}
\clearpage

\begin{figure}
\figurenum{7}
\begin{center}
\includegraphics[scale=0.7]{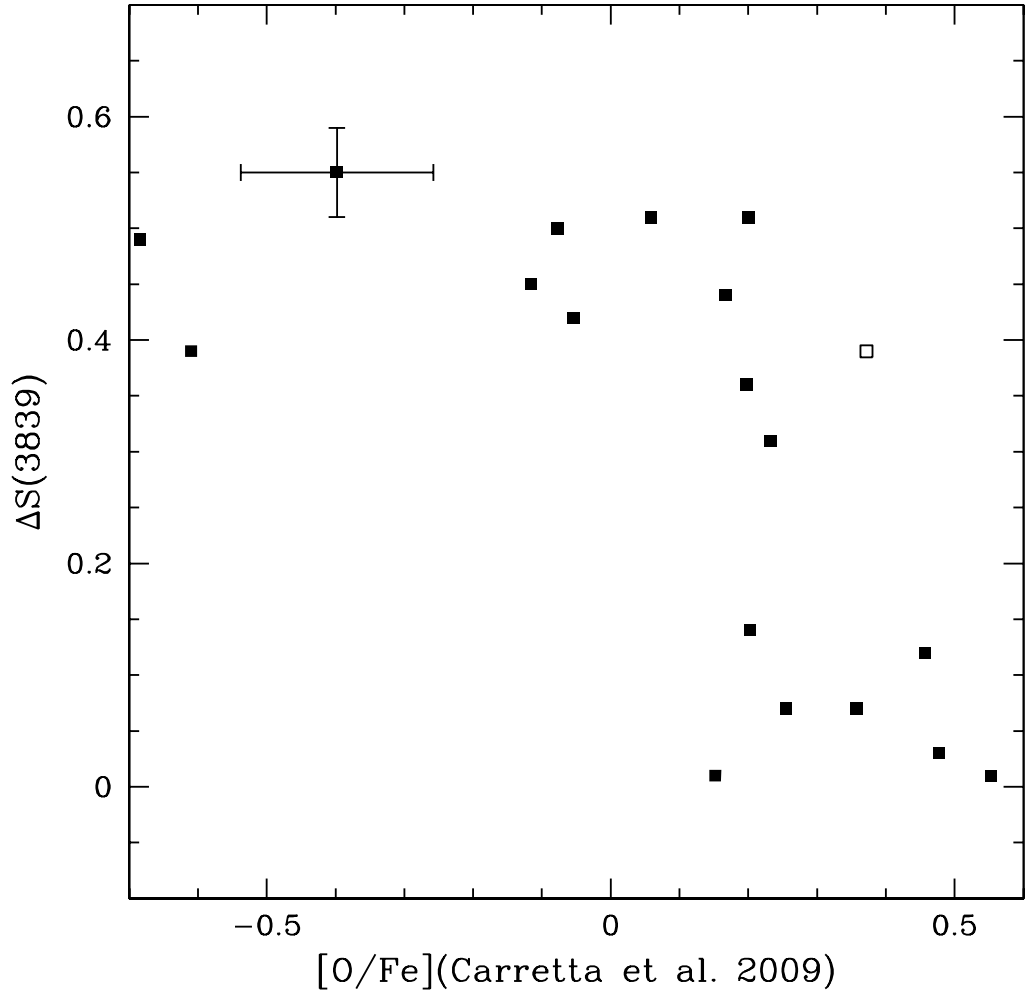}
\caption{The $\lambda$3883 CN residual $\Delta S(3839)$ versus oxygen abundance
from Carretta et al. (2009) as listed in column 9 of Table 1. Filled and open 
squares denote RGB and AGB stars respectively (see Fig.~1). Error bars of 
length $\pm 0.04$ in $\Delta S(3839)$ and $\pm 0.14$ dex in the Carretta et al.
(2009) oxygen abundance are shown for one star.}
\end{center}
\end{figure}
\clearpage

\begin{figure}
\figurenum{8}
\begin{center}
\includegraphics[scale=0.7]{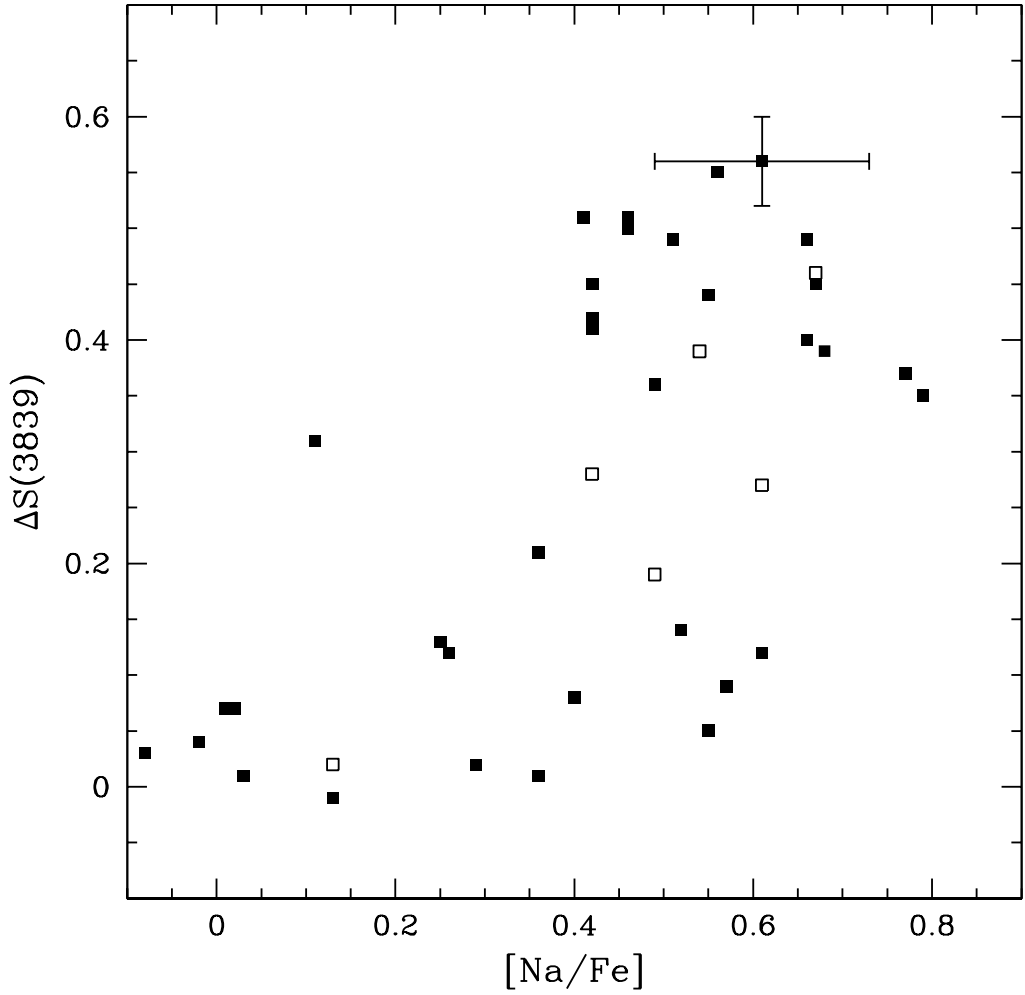}
\caption{The CN residual $\Delta S(3839)$ versus [Na/Fe] from column 7 of 
Table 1. Filled and open squares denote RGB and AGB stars respectively. A
pair of representative error bars of length $\pm 0.04$ in $\Delta S(3839)$
and $\pm 0.12$ dex in [Na/Fe] is included.}
\end{center}
\end{figure}
\clearpage

\begin{figure}
\figurenum{9}
\begin{center}
\includegraphics[scale=0.7]{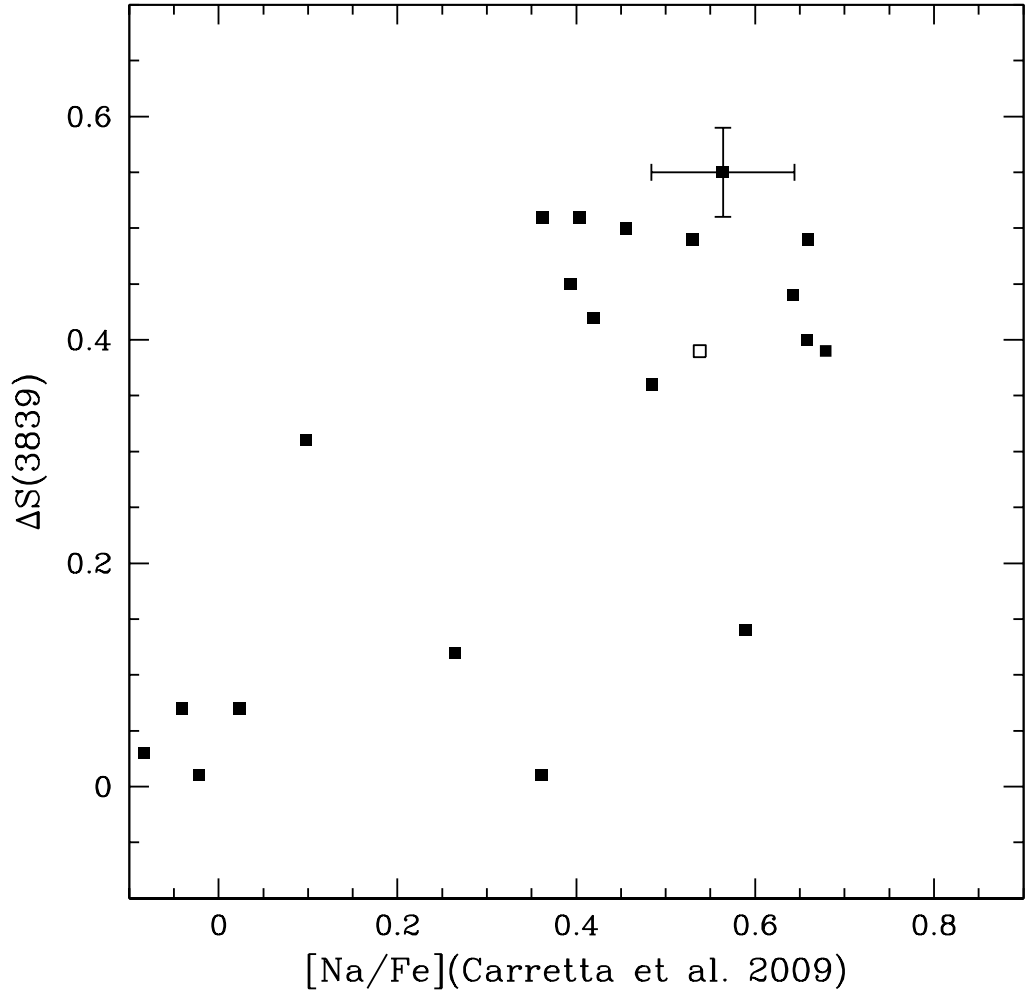}
\caption{The $\Delta S(3839)$ residual versus [Na/Fe] from Carretta et al. 
(2009) as listed in column 10 of Table 1. Filled and open squares denote RGB 
and AGB stars respectively. Typical error bars of length $\pm 0.04$ in 
$\Delta S(3839)$ and $\pm 0.08$ dex in the Carretta et al. (2009) sodium 
abundance are shown for one star.}   
\end{center}
\end{figure}
\clearpage

\begin{deluxetable}{lcccrrrcrr}
\tabletypesize{\scriptsize}
\tablecolumns{10}
\tablewidth{0pt}
\tablenum{1}
\tablecaption{Data for Evolved Giant Stars in Messier 5} 
\tablehead{
\colhead{Star}              &        
\colhead{$M_V$}             &                            
\colhead{$(B-V)_0$}         &  
\colhead{$S(3839)$}         &      
\colhead{$\Delta S(3839)$}  & 
\colhead{[O/Fe]\tablenotemark{a}}           &    
\colhead{[Na/Fe]\tablenotemark{a}}          &       
\colhead{Class}                             &
\colhead{[O/Fe](C09)\tablenotemark{b}}      &
\colhead{[Na/Fe](C09)\tablenotemark{b}}      \\
\colhead{(1)}              &       
\colhead{(2)}              &  
\colhead{(3)}              & 
\colhead{(4)}              &   
\colhead{(5)}              &  
\colhead{(6)}              &  
\colhead{(7)}              & 
\colhead{(8)}              &
\colhead{(9)}              &   
\colhead{(10)} 
}
\startdata
 I-2      &  --0.67  &  1.02  &  0.51  &   0.44  &   0.15  &   0.55  &   RGB  &   0.167  &   0.642  \\  
 I-4      &  --1.09  &  1.09  &  0.14  &   0.03  &   0.48  & --0.08  &   RGB  &   0.477  & --0.084  \\  
 I-14     &  --1.46  &  1.20  &  0.56  &   0.41  &   0.13  &   0.42  &   RGB  &   .....  &   .....  \\ 
 I-25     &  --0.88  &  1.04  &  0.45  &   0.36  &   0.20  &   0.49  &   RGB  &   0.197  &   0.485  \\  
 I-39     &  --1.36  &  1.19  &  0.78  &   0.64  &   ....  &   ....  &   RGB  &   .....  &   .....  \\
 I-50     &  --0.66  &  1.00  &  0.56\tablenotemark{c}  &   0.49  & --0.51  &   0.66  &   RGB  & --0.685  &   0.659  \\
 I-55     &  --0.84  &  0.90  &  0.27  &   0.19  & --0.33  &   0.49  &   AGB  &   .....  &   .....  \\
 I-58     &  --1.23  &  1.12  &  0.57  &   0.45  & --0.14  &   0.42  &   RGB  & --0.115  &   0.393  \\ 
 I-61     &  --1.11  &  1.11  &  0.15  &   0.04  &   0.41  & --0.02  &   RGB  &   .....  &   .....  \\
 I-68     &  --1.99  &  1.39  &  0.25  &   0.05  &   0.31  &   0.55  &   RGB  &   .....  &   .....  \\
 I-71     &  --1.38  &  1.16  &  0.49\tablenotemark{c}  &   0.35  & --0.21  &   0.79  &   RGB  &   .....  &   .....  \\
 II-39    &  --0.07  &  0.92  &  0.02  &   0.01  &   0.15  &   0.36  &   RGB  &   0.152  &   0.361  \\
 II-50    &  --0.59  &  1.00  &  0.37\tablenotemark{c}  &   0.31  &   0.22  &   0.11  &   RGB  &   0.232  &   0.098  \\
 II-59    &  --1.14  &  1.09  &  0.67  &   0.56  & --0.38  &   0.61  &   RGB  &   .....  &   .....  \\ 
 II-61    &  --1.00  &  0.99  &  0.36  &   0.26  &   ....  &   ....  &   AGB  &   .....  &   .....  \\
 II-74    &  --0.69  &  1.01  &  0.56  &   0.49  & --0.23  &   0.51  &   RGB  &   .....  &   0.530  \\
 II-85    &  --2.10  &  1.44  &  0.33  &   0.12  &   0.20  &   0.61  &   RGB  &   .....  &   .....  \\ 
 II-86    &  --1.07  &  0.99  &  0.32  &   0.21  &   ....  &   ....  &   AGB  &   .....  &   .....  \\
 III-3    &  --2.02  &  1.38  &  0.19  & --0.01  &   0.32  &   0.13  &   RGB  &   .....  &   .....  \\ 
 III-36   &  --1.65  &  1.26  &  0.31  &   0.14  &   0.22  &   0.52  &   RGB  &   0.203  &   0.589  \\
 III-50   &  --1.55  &  1.12  &  0.61  &   0.46  & --0.11  &   0.67  &   AGB  &   .....  &   .....  \\ 
 III-52   &  --0.42  &  1.01  &  0.11\tablenotemark{c}  &   0.07  &   0.37  &   0.01  &   RGB  &   0.357  & --0.041  \\
 III-53   &  --0.93  &  0.92  &  0.36  &   0.27  & --0.43  &   0.61  &   AGB  &   .....  &   .....  \\
 III-59   &  --0.53  &  0.99  &  0.56\tablenotemark{c}  &   0.51  &   0.03  &   0.41  &   RGB  &   0.059  &   0.362  \\
 III-67   &  --0.97  &  1.07  &  0.57  &   0.47  &   ....  &   ....  &   RGB  &   .....  &   .....  \\
 III-78   &  --1.83  &  1.32  &  0.19\tablenotemark{c}  &   0.01  &   0.51  &   0.03  &   RGB  &   0.553  & --0.022  \\
 III-94   &  --1.63  &  1.25  &  0.56  &   0.40  &   ....  &   0.66  &   RGB  &   .....  &   0.658  \\
 III-96   &  --1.60  &  1.23  &  0.53\tablenotemark{c}  &   0.37  &   0.04  &   0.77  &   RGB  &   .....  &   .....  \\
 III-99   &  --1.69  &  1.25  &  0.56  &   0.39  & --0.61  &   0.68  &   RGB  & --0.610  &   0.679  \\
 III-122  &  --2.16  &  1.45  &  0.31  &   0.09  & --0.06  &   0.57  &   RGB  &   .....  &   .....  \\
 IV-4     &  --0.42  &  0.98  &  0.17\tablenotemark{c}  &   0.13  &   0.25  &   0.25  &   RGB  &   .....  &   .....  \\
 IV-12    &  --0.98  &  1.08  &  0.60  &   0.50  & --0.08  &   0.46  &   RGB  & --0.078  &   0.456  \\
 IV-19    &  --1.85  &  1.32  &  0.30  &   0.12  &   0.40  &   0.26  &   RGB  &   0.457  &   0.264  \\
 IV-26    &  --0.90  &  0.93  &  0.11  &   0.02  &   0.29  &   0.13  &   AGB  &   .....  &   .....  \\
 IV-30    &  --0.97  &  0.97  &  0.38  &   0.28  &   0.12  &   0.42  &   AGB  &   .....  &   .....  \\
 IV-34    &  --1.43  &  1.18  &  0.35  &   0.21  &   0.13  &   0.36  &   RGB  &   .....  &   .....  \\
 IV-36    &  --0.51  &  0.99  &  0.56\tablenotemark{c}  &   0.51  &   0.15  &   0.46  &   RGB  &   0.200  &   0.404  \\
 IV-47    &  --2.06  &  1.39  &  0.29  &   0.08  &   0.14  &   0.40  &   RGB  &   .....  &   .....  \\
 IV-49    &  --1.32  &  1.16  &  0.68  &   0.55  & --0.40  &   0.56  &   RGB  & --0.398  &   0.564  \\
 IV-56    &  --1.28  &  1.14  &  0.55  &   0.42  & --0.05  &   0.42  &   RGB  & --0.054  &   0.419  \\
 IV-59    &  --1.80  &  1.26  &  0.57\tablenotemark{c}  &   0.39  &   0.37  &   0.54  &   AGB  &   0.372  &   0.538  \\
 IV-72    &  --1.60  &  1.26  &  0.18\tablenotemark{c}  &   0.02  &   0.31  &   0.29  &   RGB  &   .....  &   .....  \\
 IV-74    &  --0.97  &  1.07  &  0.55  &   0.45  &   ....  &   0.67  &   RGB  &   .....  &   .....  \\
 IV-82    &  --1.23  &  1.13  &  0.19  &   0.07  &   0.26  &   0.02  &   RGB  &   0.255  &   0.023  \\
 S344     &  --0.54  &  0.76  &  0.31  &   0.26  &   ....  &   ....  &   AGB  &   .....  &   .....  \\
 S445     &  --0.92  &  0.93  &  0.34  &   0.25  &   ....  &   ....  &   AGB  &   .....  &   .....  \\
\enddata
\tablenotetext{a}{Merged values from the data of Ivans et al. (2001) and Carretta et al. (2009).}
\tablenotetext{b}{Values from Carretta et al. (2009).}
\tablenotetext{c}{Based on two or three values from the literature.}

\end{deluxetable}

\clearpage

\end{document}